# THE NATIONAL INTANGIBLE RESOURCES AND THEIR IMPORTANCE IN THE CURRENT KNOWLEDGE-BASED ECONOMY

Camelia, OPREAN-STAN[1], Sebastian, STAN[2] and Antonio, PELE
[1]"Lucian Blaga" University of Sibiu, camelia.oprean@ulbsibiu.ro
[2]"Nicolae Balcescu" Land Forces Academy Sibiu, sebastian.stan@armyacademy.ro
Pontifical Catholic University of Rio de Janeiro, apele@puc-rio.br

**ABSTRACT:** In this article, models for assessing national intangible resources are analysed through a lecture in the literature, and the best-known evaluation methods are categorized into academic models and models of international organizations, with the most important differences being identified. The European Innovation Scoreboard (EIS) and the World Economic Forum annual reports on Global Competitiveness were considered to assess Romania's position in the international context in terms of intangible assets. Despite the importance of intangible resources at national level and the fact that they are an important factor in determining economic growth in the current knowledge-based economy, this article concludes that Romania's position in the international context regarding intangible assets is very weak, with many weak points in research and innovation performance compared to other EU Member States. Therefore, there is a need in our country to re-evaluate the areas where all efforts need to be focused to stimulate innovation performance, to properly manage national intangible resources, a crucial process for improving the quality of life.

## 1. INTRODUCTION: LITERATURE REVIEW

The research of intangible resources was originally developed at microeconomic level in the mid-1990s. A pioneering method for measuring an organization's intangible resources was implemented by the Swedish company Skandia [1]. As a result of specific analyses of organizations-level intangibles, several researchers have expanded this perspective and tried to measure intangible resources at nations and regions level [2]. Thus, national reports of intellectual capital have been published in several countries (e.g. in Israel, Poland and Luxembourg), and several initiatives have been launched to assess intellectual capital at national level (e.g. Sweden and Denmark) [3]. Examples at regional level for measuring intellectual capital include the National Intellectual Capital Index (NICI) for the Arab region [4] and an initiative for the Pacific Islands. Since the beginning of the millennium, intellectual capital research has advanced at a macroeconomic level. There are several variants of intangible resource evaluation models at national level, but their basics are broadly similar to the classification initially developed by Edvinsson and Malone (1997) for organizations.

Intellectual capital and the competitiveness of nations are strongly related, both being the result of knowledge within countries. Similarly, national wealth, competitiveness and national intellectual capital are some of the nation's most important goals. Various studies have shown that these objectives are strongly and directly interconnected, thus being able to create great synergies for countries [5].

Malhotra [6] has defined the knowledge within a territory as intangible resources having effects on national growth. Bontis [4] has pointed out that hidden values lie in individuals, businesses, institutions, communities and regions, and that proper management increases national wealth and economic success. Thus, the measurement and management of intangible resources improve the adoption of public policies and the use of good practices, supporting the creation of new and better investment programs, together with appropriate programs to promote development. Moreover, in their work [7], the authors tried to identify whether economic freedom, knowledge economy and global competitiveness are directly related.

Although intellectual capital is recognized at macroeconomic level as a very important factor in the determination of national wealth, this fact became evident only in the 1990s [8], with the increasing concern of intangible decision makers, not only for adaptation to the new context, but also for the relevance of intellectual capital in future national performance. The first work on macroeconomic-level intellectual capital was "Welfare and Security" by Caroline Stenfelt-Dunn, in 1996 [9]. A few years earlier, this topic was also discussed during the meeting "Managing the IC of the Nation, Managing Knowledge Assets into the 21st Century," held in 1987 in the United States of America [10]. Later, Edvinsson presented his work based on Skandia Navigator.

The development of assessment models for intangible resource at macroeconomic level followed different paths, taking into account various principles. Several studies on national intellectual capital (NIC) have replicated measurements at microeconomic level [11, 12]. Moving from the firm to the national level is based on the idea that intangible resources are just as important for countries' productivity and competitiveness as they are for firms. However, the complexity of intellectual capital assessment makes transplantation of micro-models at national level impossible [13], since the assessment of intangible resources for countries is more difficult than for firms [14].

In recent years, national intellectual capital research has generated several measurement tools to capture its impact at national level [15]. Although there is not yet a recognized macro model or widely accepted methodology to assess this intangible capital, studies of national intellectual capital and its economic impact are used as diagnostic tools to compare and analyse national development strategies [16, 17, 18].

Many studies have assessed intangible value at country level as a way to improve economic performance. Among country-level studies are included: Australia by Gans and Stern [19] and Gans and Hayes [20], Finland by Stahle and Poyhonene [21], Israel by Pasher and Shachar [22], Luxemburg by Alexander [23], Spain (Madrid) by Pomeda et al. [24], Sweden reported by Rembe [25], the Arab region by Bontis [26], EU countries by Bounfour [27] and Weziak [28] and the Nordic countries by Lin and Edvinsson [29]. Another important work consists in the evaluation of 40 countries by Lin and Edvinsson [30] care which includes a dynamic analysis for several years.

Currently, there are different models to measure intangible resources at country level, the results of which tend to converge. In addition, more attention is paid to comparative studies, especially to those reporting composite index patterns. Comparative assessments have shown a close relationship with economic performance, allowing for a better understanding of the causes of growth in the age of knowledge. Moreover, these assessments also explain the economic crises in countries such as Greece, Portugal, Italy and Spain [31], and could help to avoid them, as the assessment of intellectual capital provides information on the main intangibles that support economic growth.

## 2. ASSESSMENT OF ROMANIA'S POSITION IN INTERNATIONAL CONTEXT REGARDING INTANGIBLE RESOURCES

The European Innovation Scoreboard (EIS) and reports issued annually by the World Economic Forum on Global Competitiveness were considered to assess Romania's position in the international context in terms of intangible assets.

The annual European Innovation Scoreboard provides a benchmarking of EU Member States' research and innovation performance, while identifying their strengths and weaknesses in research and innovation systems. This picture helps Member States to assess areas where they need to focus their efforts to boost innovation performance

For EIS 2017, 16th edition, for the first time since the introduction of EIS in 2001, the measurement framework has been significantly revised. The new EIS measurement framework distinguishes between four main types of indicators and ten innovation dimensions, captured through 27 different indicators, as follows [32]:

• Framework conditions capture the main creators of the company's innovative external performance and cover three innovation dimensions: human resources, attractive research systems, and an innovation-friendly environment.
• Investments show public and private investments in research and innovation and cover two dimensions: funding and support; firms' investments.
• Innovation activities assess innovation efforts at firm level, grouped in three innovation dimensions: innovators, links and intellectual assets.
• Impacts analyse the effects of firms' innovation activities in two innovation dimensions: the impact on labour force and sales effects.

The performance of the national innovation systems of EU member countries is measured by the Summary Innovation Index, which is a composite indicator obtained by taking into consideration a weighted average of the 27 indicators. Figure 1 shows the situation for this index for all EU Member States. Thus, based on this year's results, Member States fall into four performance groups:

• The first group, of innovation leaders includes the Member States where performance exceeds the EU average by more than 20%. Leaders in innovation are Denmark, Finland, Germany, the Netherlands, Sweden and the United Kingdom.
• The second group, of strong innovators, includes member states with a performance between 90% and 120% of the EU average. Austria, Belgium, France, Ireland, Luxembourg and Slovenia are strong innovators.
• The third group, of moderate innovators includes Member States where performance ranges between 50% and 90% of the EU average. Croatia, Cyprus, Czech Republic, Estonia, Greece, Hungary, Italy, Latvia, Lithuania, Malta, Poland, Portugal, Slovakia and Spain belong to this group.
• The fourth group, of modest innovators includes Member States showing a performance level less than 50% of the EU average. This group includes Bulgaria and Romania.

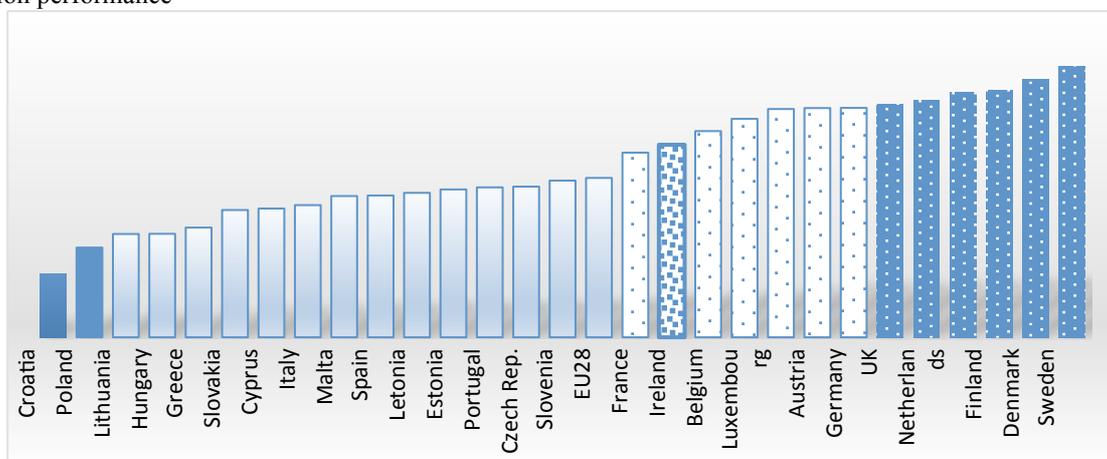

**Figure 1.** Performance of innovation systems of EU member countries in 2017. Classification according to Summary Innovation Index.

*Source: European Innovation Scoreboard 2017, 16th edition*

Romania is a modest innovator, unfortunately occupying the last position in this ranking in 2017. Compared to the European Union average (see Figure 2), the dimensions of the innovation

46

system where our country is located are related to:

- the innovation-friendly environment (with indicators such as: broadband penetration – the penetration rate of the internet, respectively the number of firms with contracts for speed internet; opportunity-driven entrepreneurship – opportunity-based entrepreneurship is a motivational index, calculated as a ratio between the share of people involved in improvement-based entrepreneurship and the share of people involved in need-based entrepreneurship);
- the innovation-friendly environment (with indicators such as: broadband penetration – the penetration rate of the internet, respectively the number of firms with contracts for speed internet; opportunity-driven entrepreneurship – opportunity-based entrepreneurship is a motivational index, calculated as a ratio between the share of people involved in improvement-based entrepreneurship and the share of people involved in need-based entrepreneurship);
- the impact of sales (with indicators such as: exports of medium and high technology products; exports of intensive knowledge services; sales of new innovative products for the market and for firms);
- human resources (with indicators such as: new PhD graduates, the 25-34-year-old population with higher education; lifelong learning, respectively the percentage of the 25-64-year-old population participating in lifelong learning).

Relative weaknesses refer to the following dimensions:

- innovators (with indicators such as: SMEs with product or process innovations, SMEs with marketing or organizational innovations, Innovative in-house SMEs);
- firms' investments (with indicators such as: research and development expenses in the business sector, innovation expenses not intended for research and development, enterprises providing training courses for developing digital skills of their staff);
- funding and support (research and development expenses in the public sector venture capital expenses).

Significant differences with the other states consist in the fact that our country has a higher share of employment in agriculture and mining; a lower share of employment in high tech, services and public administration; a larger share of foreign-controlled enterprises; a smaller number of businesses spending on cutting-edge research and development, while the value of these expenses declining on average; a declining and negative population growth rate and a decrease in population density.

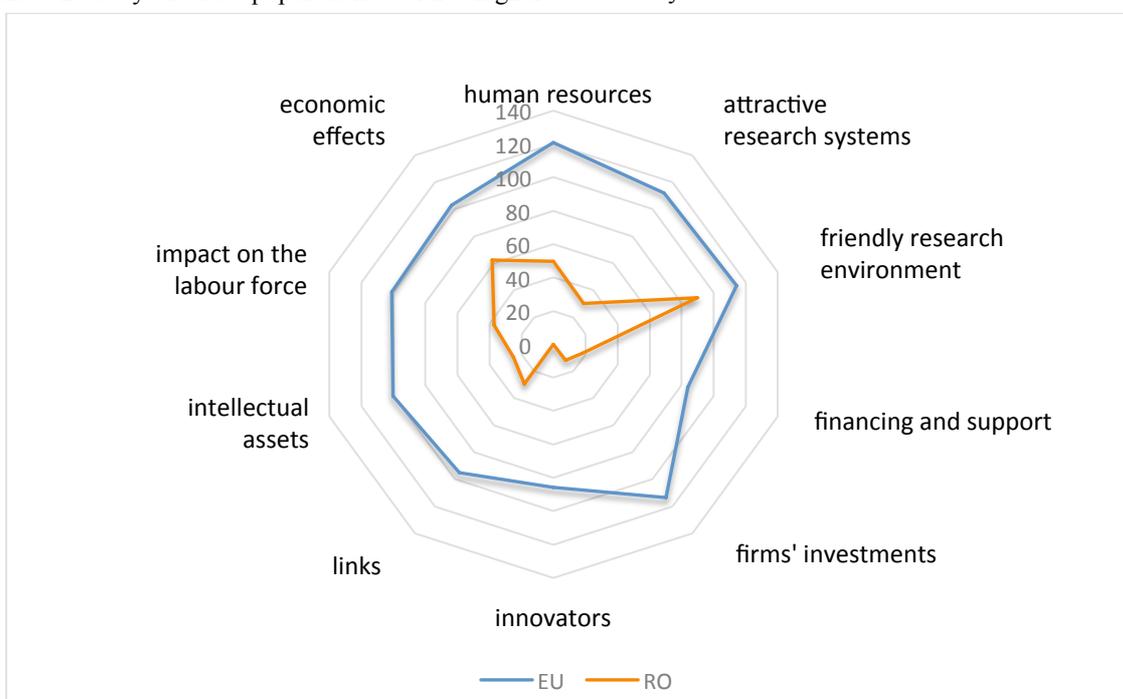

**Figure 2.** Dimensions of the innovation system in Romania compared to the European Union average.

*Source: own processing*

Romania's position in the European context can also be traced in the context presented by the World Economic Forum, by analysing the Reports issued annually by this institution on Global Competitiveness. Global Competitiveness Index (GCI), which has been developed since 2005, is based on 114 indicators incorporating various important aspects to ensure long-term productivity and prosperity for nations. These indicators are grouped into 12 pillars: institutions, infrastructure, macroeconomic environment, health and primary education, higher education and training, goods market efficiency, labour market efficiency, financial market development, technological readiness, market size, business sophistication and innovation.

The situation of Romania mirrored by the values of these pillars, extracted from the Global Competitiveness Report 2017-2018, is presented in Figure 3. All variables are expressed on a value scale from 1, the minimum value to 7, the maximum, optimal value.

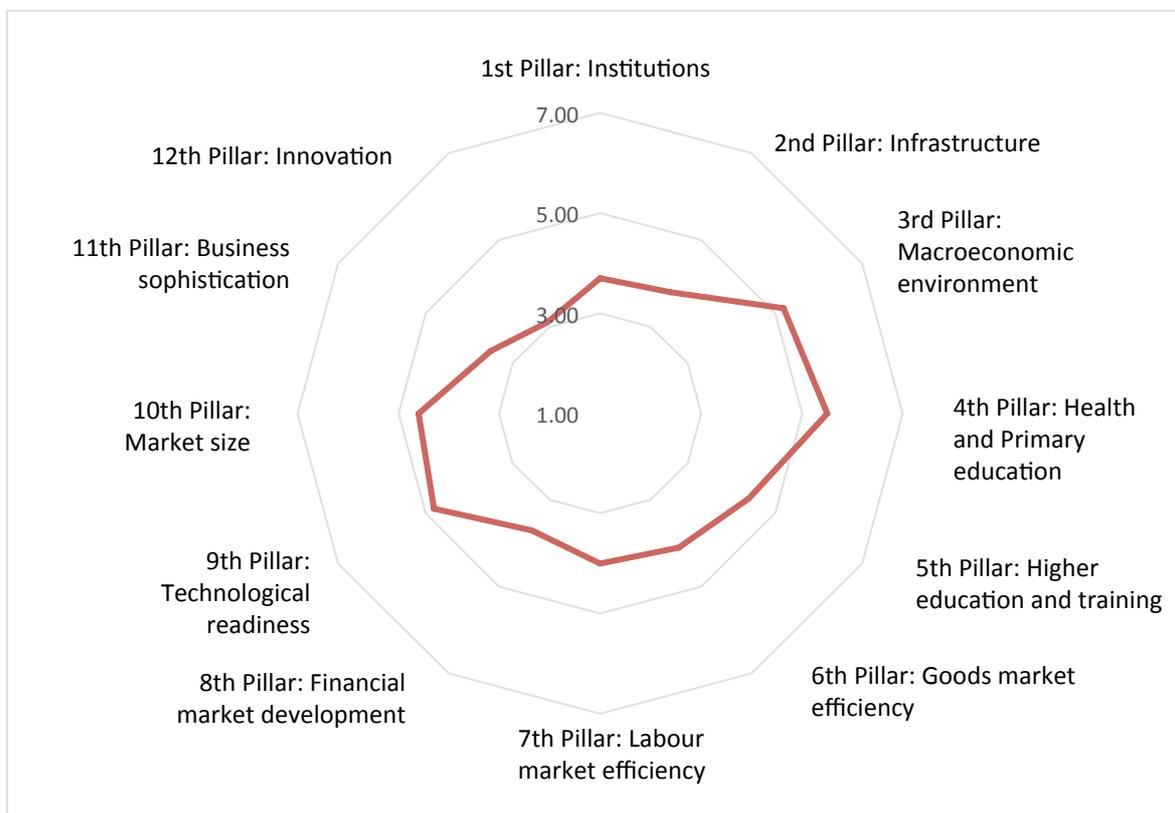

**Figure 3.** Situation of Romania exemplified by the values of global competitiveness pillars.

*Source: own processing, using data provided by the Global Competitiveness Report, World Economic Forum, period 2017-2018*

It is noted that in terms of the two pillars characteristic of performance in innovation, 11th pillar – business sophistication and 12th pillar – innovation, our country achieved the lowest values (3.5 and 3.1 respectively) compared to the other pillars. In order to better understand the aspects where our country is unsatisfactory, it is advisable to present their content. Thus, business sophistication refers to two elements that are closely related: the quality of a country's global business networks and the quality of individual firms' operations and strategies. These factors are particularly important for countries at an advanced stage of development when, to a large extent, the basic sources of productivity improvements have been exhausted. The quality of a country's business networks and supporting industries, as measured by the quantity and quality of local suppliers and the extent of their interaction, is important for several reasons. When firms and suppliers from a particular sector are interconnected in geographically close groups, called clusters, efficiency is enhanced, greater opportunities for innovation in processes and products are created, and barriers to entry for new firms are reduced. On the other hand, innovation is particularly important for knowledge-based economies. In these economies, firms need to design and develop cutting edge products and processes to maintain a competitive advantage and move towards higher value generating activities. This development requires a favourable environment for innovative activity, supported by both the public sector and the private sector. In particular, this means sufficient investment in research and development, particularly in the private sector; the presence of high-quality scientific research institutions that can generate the basic knowledge required to build new technologies; extensive collaboration in the field of technological research and development between universities and industry; protection of intellectual property. Unfortunately, Romania's situation is inappropriate in all these aspects, occupying 116th position out of 137 countries in Business Sophistication and 96th position out of 137 countries in Innovation.

## 3. CONCLUSIONS

As a conclusion, despite the importance of intangible resources at national level and the fact that they are an important factor in determining the economic growth in the current knowledge-based economy, we appreciate that Romania's position in the international context regarding intangible assets is very weak, there being many weaknesses in research and innovation performance compared to other EU Member States.

## REFERENCES

1. Edvinsson, L., Malone, M., *Intellectual Capital: Realising Your Company's True Value by Finding its Hidden Brainpower*, Harper Collins, New York, NY, (1997).
2. Ståhle Pirjo, Ståhle Sten, Lin Carol Y.Y., Intangibles and national economic wealth – a new perspective on how they are linked, *Journal of Intellectual Capital*, Vol. 16 Iss 1 pp. 20 – 57, (2015).
3. Bounfour, A., The IC-dVAL approach, *Journal of Intellectual Capital*, Vol. 4 No. 3, pp. 396-413, (2003).
4. Bontis, N., National intellectual capital index: a United Nations initiative for the Arab region, *Journal of Intellectual Capital*, Vol. 5 No. 1, pp. 13-39, (2004).
5. Herciu M., Ogrean C., Wealth, Competitiveness, and Intellectual Capital–Sources for Economic Development 2015/1/1 *Procedia Economics and Finance* 27, 556-566 Elsevier, (2015).
6. Malhotra, Y., Measuring knowledge assets of a nation: knowledge systems for development, Invited Research Paper Sponsored by the United Nations Department of Economic and Social Affairs. Keynote Presentation at the *Ad Hoc Group of Experts Meeting at the United Nations Headquarters*, New York City, NY, (2003).

7. Herciu M., Ogrean C., Interrelations Between Economic Freedom, Knowledge Economy And Global Competitiveness–Comparative Analysis Romania And Eu Average, *Studies in Business and Economics* 6 (2), 46-59, (2011).
8. Lopez Ruiz V.R., Nevado Pena D., Alfaro Navarro J.L., Badea L., Grigorescu A., Voinea L., Measurement of National Non-Visible Wealth through Intellectual Capital, *Romanian Journal of Economic Forecasting*, 3/2011, pp. 200-212, (2011).
9. Edvinsson, L., Stenfelt, C., Intellectual capital of nations – for future wealth creation, *Journal of Human Resource Costing & Accounting*, Vol. 4 No. 1, pp. 21-23, (1999).
10. Labra Romilio, Sánchez M. Paloma, National intellectual capital assessment models: a literature review, *Journal of Intellectual Capital*, Vol. 14 Iss 4 pp. 582 – 607, (2013).
11. Ståhle, P., Ståhle, S., Aho, S., Value added intellectual coefficient (VAIC): a critical analysis, *Journal of Intellectual Capital*, Vol. 9 No. 4, pp. 531-551, (2011).
12. Lazuka, V., *National intellectual capital: concept and measurement*, Master programme in economic growth, innovation and spatial dynamics, Lund University, School of Economics and Management, Lund, available at: http://aspheramedia.com/wp-content/uploads/2015/07/Thesis_Volha_Lazuka_FINAL.pdf, (2012).
13. Lin, C.Y.Y., Edvinsson, L., *National Intellectual Capital, A Comparison of 40 Countries*, Springer, New York, NY, ISBN 978-1-4419-7376-1, (2011).
14. Kapyla, J., Kujansivu, P., Lonnqvist, A., National intellectual capital performance: a strategic approach, *Journal of Intellectual Capital*, Vol. 13 No. 3, pp. 343-362, (2012).
15. Salonius, H., Lonnqvist, A., Exploring the policy relevance of national intellectual capital information, *Journal of Intellectual Capital*, Vol. 13 No. 3, pp. 331-342, (2012).
16. Andriessen, D. Stam, C., Intellectual capital of the European Union, *7th McMaster World Congress on the Management of Intellectual Capital and Innovation*, Hamilton, ON, January 19-21, available at: http://213.155.109.122/files/sources/19_20111215032255185.pdf, (2005).
17. The Government of Poland, *The report on intellectual capital of Poland*, available at: http://zds.kprm.gov.pl/en/report-on-intellectual-capital-of-poland, (2008).
18. Alfaro, J., Lopez, V. and Nevado, D., An alternative to measure national intellectual capital adapted from business level, *African Journal of Business Management*, Vol. 5 No. 16, pp. 6707-6716, (2011).
19. Gans, J., Stern, S., *Assessing Australia's innovative capacity in the 21st century*, Intellectual Property Research Institute of Australia – IPRIA Melbourne, (2003).
20. Gans, J., Hayes, R., *Assessing Australia's Innovative Capacity: 2007 Update*. Centre for Ideas and the Economy, Melbourne Business School, University of Melbourne, Melbourne, available at: http://works.bepress.com/joshuagans/16, (2008).
21. Ståhle, P., Poyhonene, A., Intellectual capital and national competitiveness: a critical examination. Case Finland, paper presented at the *6th European Conference of Knowledge Management* (ECKM), University of Limerick, (2005).
22. Pasher, E., Shachar, S., *The intellectual capital of state of Israel*, in Bounfour, A., Edvinsson, L. (Eds) Intellectual Capital for Communities. Nations, Regions and Cities, Elsevier Butterworth-Heinemann, Burlington, MA, pp. 139-150, (2005).
23. Alexander, S., An intellectual capital audit of the Grand Duchy of Luxembourg, *The World Conference on Intellectual Capital for Communities*, World Bank Office, Paris, June 29-30, (2006).
24. Pomeda, J., Merino, C., Murcia, C., Vollar, L., Towards an intellectual capital report of Madrid: new insights and developments, paper presented at *The Transparent Enterprise. The Value of Intangibles*, Madrid, 25-26 November, (2002).
25. Rembe, A., *Invest in Sweden: Report 1999*, Halls Offset AB, Stockholm, (1999).
26. Bontis, N., National intellectual capital index: a United Nations initiative for the Arab region, *Journal of Intellectual Capital*, Vol. 5 No. 1, pp. 13-39, (2004).
27. Bounfour, A., The IC-dVAL approach, *Journal of Intellectual Capital*, Vol. 4 No. 3, pp. 396-413, (2003).
28. Weziak, D., *Measurement of national intellectual capital: application to EU countries*, IRISS Working Paper Series No.13, INSEAD, November Lifferdange, (2007).
29. Lin, C.Y.Y., Edvinsson, L., National intellectual capital: comparison of the Nordic countries, *Journal of Intellectual Capital*, Vol. 9 No. 4, pp. 525-545, (2008).
30. Lin, C.Y.Y., Edvinsson, L., *National Intellectual Capital, A Comparison of 40 Countries*, Springer, New York, NY, ISBN 978-1-4419-7376-1, (2011).
31. Lin, C.Y.Y., Edvinsson, L., Chen, J., Beding, T., *National Intellectual Capital and the Financial Crisis in Greece, Italy, Portugal, and Spain* (ISBN: 978-1-4614-5989-7), Springer Science and Business Media, LLC, New York, NY, (2013).
7. European Innovation Scoreboard 2017, European Commission, available at: http://ec.europa.eu/growth/industry/innovation/facts-figures/scoreboards.